\documentclass[5p,preprint,authoryear]{elsarticle}

\usepackage{todonotes}
\usepackage[utf8]{inputenc}
\usepackage{color}
\usepackage{listings}
\usepackage{tikz}
\usepackage{hyperref}
\usetikzlibrary{shapes,calc,arrows,mindmap,positioning,decorations.pathreplacing,fadings}
\usepackage{xcolor}
\usepackage{float}
\usepackage{amsmath}

\lstdefinestyle{base}{
  language=C++,
  emptylines=1,
  breaklines=true,
  basicstyle=\color{black}\ttfamily,
  keywordstyle=\color{blue}\ttfamily,
  commentstyle=\color{green!50!black}\ttfamily,
  morecomment=[l][\color{magenta}]{\#},
  moredelim=**[is][\color{blue!50!green}]{@}{@},
}
\lstdefinestyle{rightnums}{
  numbers=right
}
\lstdefinelanguage{configfile}{
	keywords={default,binname},
	keywordstyle=\color{blue}\bfseries,
	identifierstyle=\color{black},
	sensitive=false,
	comment=[l]{\#}
}
\usepackage{subcaption}
\graphicspath{{./res/}}

\usepackage{hyperref}

\journal{SoftwareX}

\begin{document}
	\begin{frontmatter}
		\title{Rambrain - a library for virtually extending physical memory}
		\author[label1,label2]{Imgrund, M.}
		\ead{imgrund@usm.uni-muenchen.de}
		\ead[url]{https://github.com/mimgrund/rambrain/}
		\author[label1,label3]{Arth, A.}
		\ead{arth@usm.uni-muenchen.de}
		\address[label1]{University Observatory Munich, Scheinerstraße 1, 81679 Munich, Germany}
		\address[label2]{Max-Planck-Institute for Radio Astronomy, Auf dem Hügel 69, 53121 Bonn, Germany}
		\address[label3]{Max-Planck-Institute for Extraterrestrial Physics, Giessenbachstrasse 1, 85748 Garching, Germany}
		
		\begin{abstract}
		We introduce Rambrain, a user space C++ library that manages memory consumption of data-intense applications. Using Rambrain one can overcommit memory beyond the size of physical memory present in the system. While there exist other more advanced techniques to solve this problem, Rambrain focusses on saving development time by providing a fast, general and easy-to-use solution. Rambrain takes care of temporarily swapping out data to disk and can handle multiples of the physical memory size present. Rambrain is thread-safe, OpenMP and MPI compatible and supports asynchronous IO. The library is designed to require minimal changes to existing programs and pose only a small overhead.
		\end{abstract}
		
		\begin{keyword}
			memory management; physical memory limitations; abstraction library; system paging; open source; MPI/OpenMP
		\end{keyword}
	\end{frontmatter}
	
	\begin{table*}[!ht]
		\centering
		\begin{tabular}{|l|p{6.5cm}|p{6.5cm}|}
			\hline
			\textbf{Nr.} & \textbf{Code metadata description} &  \\
			\hline
			C1 & Current code version & 1.1 \\
			\hline
			C2 & Permanent link to code/repository used for this code version & $https://github.com/mimgrund/rambrain$ \\
			\hline
			C3 & Legal Code License   & GPL \\
			\hline
			C4 & Code versioning system used & git \\
			\hline
			C5 & Software code languages, tools, and services used & C++, OpenMP, MPI \\
			\hline
			C6 & Compilation requirements, operating environments \& dependencies & Linux, libaio \\
			\hline
			C7 & If available Link to developer documentation/manual & $http://mimgrund.github.io/rambrain/$ \\
			\hline
			C8 & Support email for questions & arth@usm.uni-muenchen.de \\
			\hline
		\end{tabular}
		\caption{Code metadata}
		\label{} 
	\end{table*}
	
	\section{Introduction}
			\label{sec:intro}
	Facing large amounts of data, be it simulations or observation results, many astrophysicists have become part-time software engineers. As the primary target of their work focuses on producing astrophysical results, developing data analysis code is an inevitable obstacle on their way to the actual goal. In the case of the authors this goal is respectively to analyse extensive data sets of pulsar timing information \citep[based on][]{imgrund2015} and to post-process large snapshots of cosmological simulations (see Arth et al. in prep.). While typical software-engineering amounts to serialising given tasks to be executed as quickly as possible, many everyday codes evaluating data or simulation results are written to be run only a few times. In this light, the primary focus of an astrophysicist often lies on saving development time and not execution time.\\
	Writing code that processes large data sets is one of the most time consuming tasks. When developing applications that use large amounts of main memory, a single larger dataset may suffice for the system to run out of memory. The typically chosen solution to this is finding a machine with more main memory. It is obvious that this solution is only temporary when facing growing amounts of data. The sophisticated solution amounts to writing memory management functions in an optimised but specialised way for the problem at hand, so called ``out-of-core computing''. This, however, is very (development) time consuming.\\
	Alternatively, one can think of following the typical approach nowadays, which has been made possible by ongoing hardware developments, and solve the memory shortage by parallelising one's code. In addition to a common computing cluster hardware vendors increase the amount of possibilities by introducing additional components like non volatile memory (NVRAM) or memory with high bandwidth (MCDRAM). However, the task of parallelising remains and is in general non trivial to implement since a distributed memory parallelisation, for example using MPI, has to be chosen. Additionally, not every code scales properly. Thus, one might run into the issue of wasting a lot of CPU time, which has to be granted after writing computing proposals, just to fulfil memory requirements.\\
	Therefore, we introduce Rambrain, a library that facilitates quick development of applications in need of large main memory. It is built to easily integrate with existing C++ code on Linux and helps applications to swap out temporarily unneeded data to transparently access multiples of the actual physical memory available on the system.\\
	While there may exist other solutions more specific to the problem at hand showing slightly better performance, we argue that in most situations the flexibility of a fast, reliable and out-of-the-box solution is preferred to a few percent performance gain. In the following, we provide a quick review of other solutions to the problem at hand and discuss in which cases rambrain might be a superior choice.
\section{Common strategies to avoid out-of-memory errors}
	The most basic strategy to still run an application in a situation of scarce free memory is using native system swapping. Modern operating systems like Linux manage association of physical memory to various processes running at a given moment. As an application developer, you are presented a more or less consecutive virtual memory address space. It is in general not clear whether a chunk of virtual memory, a so called ``page", is residing in a physical main memory location, called a ``frame", at a given time or not. This layer of abstraction facilitates assignment of memory to a process, so that the system can overcommit physical memory and reassign virtual pages to physical frames, when desired. When free frames become scarce, the system writes out currently unused pages to secondary storage (such as hard disks) in order to free frames. When a process tries to access a non-resident page, a page fault is triggered and the page is read in from secondary storage by the memory manager of the system \citep[p.20]{artofmem} and if necessary, according frames are freed by writing the occupying pages out beforehand. While this process is efficient under normal operation, the system typically slows down to being unusable when actively consuming nearly all physical memory. Especially when multiple processes compete for the remaining space (a typical situation for a developer working and debugging), the computer is virtually unusable until the memory-intense calculation has finished. How long a system can survive in a usable state might be dependent of the type of secondary storage employed. For example a SSD may keep a system usable for a longer time than a common HDD just because of it's higher speed of reading and writing data. Inevitably, the system will be still overwhelmed by the amount of data scheduled for transfer and especially the concurrent requests due to multitasking.\\
	This swapping mechanism is also limited by the available swap space on the secondary storage. While adding more swap space with the system's on-board mechanisms\footnote{Using the system tools \emph{mkswap}/\emph{swapon} as root.} is possible, it needs super user privileges and reserves the whole swap size on the disk even if it is not used completely. Furthermore, it aggravates the situation when multiple processes are competing for memory, as more and more parts of other programs can be swapped out and need to be swapped in again in order to continue execution.\\
	Using system swapping as a mechanism for overcommitting main memory can also provoke the action of the so called ``Out-Of-Memory Killer (OOM-Killer)". As available memory becomes sparse, the system tries to keep most processes running. In order to free memory for other processes, the OOM-Killer will kill one or more processes by assigning a score correlated with importance, memory consumption, execution and idle times of the candidate process. The OOM-Killer thus can abort simulation or analysis at the very last step and protections against it are hard to find \citep[see e.g.][]{oom1}. The OOM-Killer can by now be controlled a bit finer via the \emph{/proc} file system, but shutting it off for a certain process needs administrator privileges. However, one has to keep in mind that even if one can force the own application to stay alive, the OOM-Killer can simply shut down system processes which may trigger secondary effects on the target process. To the knowledge of the authors, it is not possible to completely turn off the OOM-Killer on every system. This becomes clear when concerning the alternatives in a situation of low RAM. A call to the \verb+sbrk+-family of functions to increase heap size could possibly block indefinitely, locking the process that called for more memory. Unless any other process will free memory or terminate, the next process demanding for more heap memory will block too. The resulting cascade of blocking processes would probably have much worse consequences for system health than killing a specific process based on a reasonable metric.\\
	There exist other global kernel parameters such as kernel 'swappiness' to manipulate kernel swapping behaviour. At first glance, decreasing or increasing the amount of pre-emptive swap out of idle application's virtual memory to disk sounds like a reasonable strategy to globally keep the system efficiently in function. Tuning this parameter, however, is only useful when the amount of free physical memory is huge compared to the problem at hand. While low values of this parameter will delay starting to swap out considerably, the demand of the main application for more RAM will dominate at some point below the physical memory size.\\
	In addition, such global tweaks have to be applied system wide. While a user space solution like Rambrain can be allied to any system at hand, it requires very good corporation with system administrators to employ such a behaviour on a managed machine.\\
	The next often mentioned solution to memory and swap management is the \verb+mlock+ and \verb+mmap+ family of kernel functions.\\
	\verb+mlock+ is capable of locking address ranges for kernel swap out and can also advice the kernel to swap in ranges of memory from the swap space. While these functions can be a usable approach for real-time applications that rely on fast memory access, it in no way limits heap growth. Thinking from the perspective of 'freeing physical memory for new calculations', the functions are of very limited use, as one cannot force the operating system to write out data to swap and there is no guarantee that this will affect physical process size at all.\\
	The \verb+mmap+-family of functions is used to seamlessly map disk files to virtual address space. The file can then be manipulated as if it were resident at that virtual address space location. Combined with \verb+mlock+ calls, the user is able to finely tune which parts of a file will be resident in physical memory. There even exists an interface that can be used to track which parts of a file currently reside in physical memory. Also, the memory mapped regions are accounted for as cache, thus this memory will be swapped away preferably when system memory becomes low, which reduces the overall memory footprint of the application. However, usage of memory maps for large files effectively can be very complicated, as it may only be reasonable to open certain 'windows' into regions of the file used for swapping and the number of regions is limited by file descriptor limits. 
	Such a more controllable user-space solution is desirable, for example combining the memory mapping system calls with moderate sized swap files on the secondary storage. Memory mapping techniques are fast because they use the same paging and copy mechanisms such as system swapping, but are subject to stronger limitations than letting the system handle the paging itself.\footnote{Both the number and size of memory maps are limited by the system.} The consecutive logical address space that is handed over to the process has to be managed by the user. This means that the user has to take care of allocating multiple data structures on top of the space, a mechanism that the new/delete operators deal with in C++, normally. While handling for example a vector of fixed size structures in a memory map is simple, allocating objects of different sizes is highly non-trivial. As the system is responsible for writing out the memory mapped regions to the file on secondary storage, efficient interaction with the kernel when changing the memory-mapped region is challenging when trying to optimise this process for performance. Furthermore, a strategy deciding which contiguous region to swap out is all but clear.\\
	The authors in fact started to write a backend for the actual swapping I/O of Rambrain with memory mapped files. On the long run, it turned out to be much more complicated to synchronize the swapping behaviour of the mapped regions to gain performance without knowing the exact access pattern of the user beforehand and having only a few guarantees from the Linux kernel API. Thus, a perhaps more performant solution to a problem at hand can be implemented using these facilities, but this turns out to be a difficult encounter that will at least lead to complicated memory management code. Rambrain wants to facilitate development of memory-intensive applications and is designed to take the burden of writing exactly such code from the user. In that respect, Rambrain will not beat a custom tailored solution, but coding such a solution is a hard task in its own respect. This renders such a technique possible, but complicates robust implementation and favourable run time behaviour in highly dynamic situations.\\
	Of course, there exist already other solutions to tackle large data structures in memory, such as the STXXL \citep{SPE:SPE844} that facilitate out-of-core computation providing large standard containers in analogy to the Standard Template Library (STL). While this is a very useful idea, it has still some drawbacks imposed by it's specialised approach. Rambrain has built in class support for the full C++ standard in contrast to the limitation to POD-support of the STXXL. Rambrain provides direct access to pointers in memory and thus will pose no overhead over heap allocation once the pointers have been provided. Additionally, objects created with Rambrain can be used in association with normal STL-containers and will be swapped, too.\\
	An alternative approach, using parallel virtual file systems is also imaginable \citep[see for example][]{Tang2004}. However, this kind of approach still leaves the programmer with the burden to write IO operations himself, even if they may be encapsulated e.g. as a function.\\
	Furthermore, optimising the data flow on this level comes near to developing an out-of-core algorithm for the problem at hand that takes control over all input and output operations manually. Introductory reviews of such algorithms can be found in \cite{Toledo,Vitter}. Of course one can design a very clever way of handling input and output data to boost performance. This, however, opposes the goal to find a more generic solution that gives the developer moderate control over input and output flow while taking from him the burden of handling the input and output manually. Specialised solutions cover for example n-body codes \citep{Salmon97} or linear algebra calculations \citep{Toledo1999b,Reiley1999}.\\
	From the view of the application developer, the situation is very simple: When writing a program the developer knows what data he uses, what he will use next, and what is not needed for longer time. This information is always present directly in the source code. In the next section we will introduce the interface which communicates this information to the library.
	
	\section{Interfacing Rambrain}
	\label{sec:usage}
	In order to manage the storage needs of a C++ application, we are faced with the problem of designing an interface to tell Rambrain, which data is to be managed and when it has to be present. In this chapter we introduce this interface built to require minimal changes of existing code while at the same time providing rich convenience features when possible.
		\subsection{Basic usage}
		 \label{sec:basicusage}
		As a memory manager keeping track of data has some overhead on its own, it is only useful when the data managed is large. Rambrain can manage simple primitives, arrays, whole classes and also supports nesting of managed objects into managed classes. For a start, consider the code in listing \ref{lst:stdinit} that is initialising a two dimensional plane wave field of data type double on heap memory.
		  \begin{lstlisting}[style=base,basicstyle=\small,caption=Typical two dimensional field initialisation,label=lst:stdinit,frame=single,float,floatplacement=H]
double k_x=1.,k_y=1.;
unsigned int x_max=1024, y_max=1024;
		  
double *arr[x_max];
for (int x=0;x<x_max;++x) //allocate rows
  arr[x] = new double[y_max];
for (int x=0;x<x_max;++x){ //initialize field
  double *line = arr[x];
  double xx = x / (double) x_max;
  for (int y=0;y<y_max;++y){
    double yy = y/(double) y_max;
    line[y] = sin((xx*k_x+yy*k_y));
  }
}
//do something and delete afterwards:
for (int x=0;x<x_max;++x)
  delete arr[x]; //deallocate lines
\end{lstlisting}
\begin{lstlisting}[style=base,style=rightnums,basicstyle=\small,caption=typical two dimensional field initialisation with Rambrain,label=lst:raminit,frame=single,float,floatplacement=H]
double k_x=1.,k_y=1.;
unsigned int x_max=1024, y_max=1024;

@managedPtr<@double@>@ *arr[x_max];
for (int x=0;x<x_max;++x) //allocate rows
  arr[x] = new @managedPtr<@double@>(@y_max@)@;
for (int x=0;x<x_max;++x){ //initialize field
  @adhereTo<double> glue(arr[x])@;
  double *line = @glue@;
  double xx = x / (double) x_max;
  for (int y=0;y<y_max;++y){
    double yy = y/(double) y_max;
    line[y] = sin((xx*k_x+yy*k_y));
  }
}
//do something and delete afterwards:
for (int x=0;x<x_max;++x)
  delete arr[x]; //deallocate lines
\end{lstlisting}
We allocate an array of pointers to the respective field rows in line 4, allocate the actual rows in line 6, and set up a plane wave over all field values in lines 7 to 14. Some calculations are executed prior to the deallocation of the rows in line 17. \\
If we assume now that \verb+y_max+ and \verb+x_max+ take large values, the allocated doubles will consume a non-negligible amount of RAM, passing a gigabyte at roughly $11600^2$ elements. Thus, the developer would have to swap out elements if he seeks to avoid system-swapping to occur, to ensure that the program does not run out of physical memory. Manual implementation inserts many lines of code when allocating memory and around line 8. Alternatively, the user would write his own memory manager version calling functions to load and unload data. When several objects are needed at once, loading and unloading become the dominant part of the code.\\
Furthermore the additional lines start to obfuscate algorithmic code structure. The nested \verb+for+-loops as well as the essential initialisation done will be difficult to spot. Minimal changes to this passage of code will allocate the arrays so that Rambrain is aware of them and dynamically loads and unloads the lines if needed, as can be seen in listing \ref{lst:raminit}.\\
The overall structure is minimally changed. Up to adding line 8 we only wrap data objects. We introduce two template classes here, \verb+managedPtr<>+ and  \verb+adhereTo<>+ to emplace Rambrain. When using Rambrain in a minimal way, these two classes will be the only ones actively referenced by the developer.\\
The first class, \verb+managedPtr<>+, replaces allocation and deallocation by Rambrain wrappers. This replacement is necessary to hide away the pointer to the actual data in logical memory, as the element may or may not be present when the user dereferences that pointer.\\
Consequently, we need a way to give back access to the data. This is done by \verb+adhereTo<>+ which states its meaning in camel-case: This objects adheres to the data. While the respective \verb+adhereTo<>+ object exists according to scoping rules, it is guaranteed that the user can fetch a valid pointer to the data by assigning the \verb+adhereTo<>+ object to the pointer, as is done in line 9. In the following, we will also refer to this as ``pulling the pointer".\\
The scoping relieves the user from the need to explicitly state that the data is no longer used for the moment. While the corresponding \verb+adhereTo<>+ object exists, the pointer to the data remains valid. When this ``glue'' to a \verb+managedPtr<>+ is deleted, for example by going out of scope, the object may be swapped out to disk in order to free space in physical memory for other objects, if needed.\\
This already concludes what a developer needs to know about Rambrain to write his own code using the library in the most basic fashion.
		\subsection{Advanced usage}
		 \label{sec:advacedusage}
Currently, Rambrain is, amongst others, equipped with the following advanced features that give more detailed control or convenience. The line numbers given refer to the code examples in listing \ref{lst:advandecinline}. The advanced features show that the interface is both minimalistic and powerful enough to facilitate development with Rambrain.
\begin{lstlisting}[style=base,style=rightnums,basicstyle=\small,frame=single,label=lst:advandecinline,caption=Advanced features,float,floatplacement=H,stepnumber=1]
managedPtr<double> a1; //single element
managedPtr<double> a2(5); //array of five elements
managedPtr<double> a3(5,1.); //five elements, all set to 1.
managedPtr<double,2> a1(5,5,0); //two dim., vals set to 0.

class B { public:
  B(); B(double &a, double &b);
  ~B();
  void someFunction();
  managedPtr<double> data; } //Class with ctors/dtor

managedPtr<B> b1; //single element, default constructor
managedPtr<B> b2(1) //single element, default constructor
managedPtr<B> b2(1,a,b); //single element, param. ctor
managedPtr<B> b2(5,a,b); //5 elements, parametrised ctor

adhereTo<double> glue1(a1);//Load right away
adhereTo<double> glue2(a2,false); // Load when used
const adhereTo<double> glue3(a3); // Access const

double *c1=glue1;
double *c2=glue2; //If not present, will be fetched here
const double *c3 = glue3;

//= adhereTo<double> a1_glue(a1); double* a1data = a1_glue;
ADHERETOLOC(double, a1, a1data); 

void B::someFunction(){
  ADHERETO(double,data); //shadows member B::data
  data[0] = 42.; }

//MT: Do not fail if too much memory is requested:
managedMemory::defaultManager->setOutOfSwapIsFatal(false);
//MT: Avoid deadlock when needing multiple data at once:
double *c5,*c6;
adhereTo<double> c5_glue(a1),c6_glue(a2);
{LISTOFINGREDIENTS
  c5 = c5_glue;
  c6 = c6_glue; }
\end{lstlisting}
\begin{itemize}
\item {\bf Allocation of simple datatypes.} The user may allocate a single object or multiple objects at once, passing an initial value. Also multidimensional arrays are supported, that will be collapsed to an array of \verb+managedPtr<>+s of the size of the last dimension. (lines 1-4)
\item {\bf Class allocation.} Class objects may have nested \verb+managedPtr<>+s which can be swapped out independently of the class object. Rambrain supports parametrised as well as default constructors. Destructors will be called in the correct sequence. Furthermore, the member hierarchy can be tracked. Finally, Rambrain will ensure correct deallocation of the object. As some or all parts of it may have been swapped out, this is a non-trivial task. The code supports array initialisation on classes, too. (lines 6-15)
\item {\bf Different kinds of loading stages.} The user may explicitly state whether to load objects immediately or delay actual loading until the first pointer is being pulled from the \verb+adhereTo<>+ object. \\ Rambrain can profit from \verb+const+-accessing the data. In case of the object having been swapped out already, the swap file copy is not changed and reused and thus another write-out is not necessary. If the developer requests write access, the object has to be rewritten to the file system for a swap-out. Therefore, when only reading data, using $const$-pointers is highly encouraged as will be seen in section \ref{sec:const}. (lines 17-23)
\item {\bf Convenience macros.} When adhering to an object and pulling a pointer should happen in the same slot, we provide convenience macros that create the \verb+adhereTo<>+-object together with pulling a pointer in a single line. For class members this may happen shadowing a parameter. In this case, the resulting code reads as if the class would contain an unmanaged array of the same name. Of course, $const$-versions of these macros exist, too. (lines 25-30)
\item {\bf Multithreading options.} When using Rambrain in a single threaded context, Rambrain throws an exception when the user tries to pull pointers referencing more data than the physical memory limit at once. This can be disabled by a function call to enable over-commitment in multithreaded situations. In this case, pulling a pointer that would violate the limits blocks until enough RAM has become available by other threads destroying their \verb+adhereTo<>+s. (line 33). However, this can potentially introduce a deadlock. Take for example a couple of threads that need two pointers each to start their calculation. Assume only half or less of these \verb+managedPtr<>+s fit into RAM. In this case, all or some threads may have requested the first of the needed two pointer in parallel. Since Rambrain cannot free pulled pointers while the respective \verb+adhereTo<>+s in scope exist, it blocks all threads and waits for memory to become available to swap-out. This, however, will never happen, as all threads are waiting and no thread is eventually finishing to unlock data for swapping. To circumvent this situation, the user may use a globally locking scope conveniently provided by Rambrain (lines 37-39). It is however highly encouraged not to over-commit memory also in multi-threaded situations as performance may drop by this forced serialisation.
\end{itemize}
		\subsection{Design considerations for user code}
		  \label{sec:designusage}
Having introduced the basic usage style of the library, let us evaluate the impact of using Rambrain on code design. While the syntax suggests that there would be nothing to keep in mind, a few limits and caveats apply nevertheless.
\subsubsection{Maximum problem size}
Rambrain's physical memory usage is limited to a certain amount the \verb+managedPtr<>+s may consume.\footnote{Currently we do not track the overhead imposed by the usage of Rambrain, as well as other heap allocations. This is planned for a future release.} As Rambrain cannot use the native OS paging mechanisms, it is bound to the memory limits set by the user. Consequently, the set of currently existing \verb+adhereTo<>+s\footnote{Explicit delayed loading can be emplaced to limit this to the set of adhereTo$<>$s that a pointer was pulled from.} marks data as in-use and determines what cannot be swapped out. Additional managed pointers may only consume the remaining free memory. Thus, Rambrain will be unable to manage problems that demand the simultaneous use of more data than this limit. The code has to be written in a way that the maximum simultaneously accessed data amounts to less bytes than the limit. This usually is the case anyway as algorithms are being formulated in a local way on the data.
\subsubsection{Data structures}
The size of the simultaneously used data structures relates to the way of solving a problem. A matrix operation, for example, can typically be formulated on various matrix representations such as rows, columns, sparse single elements or smaller submatrices. To gain something from managing such a subobject, the user has to take care that the payload per managed pointer is large enough, so that the overhead of managing the data becomes small. We propose allocating smaller structures via traditional mechanisms and leaving the data-intense elements to Rambrain. If however a \verb+managedPtr<>+ is chosen, it is vital to keep in mind that this block of data can only be swapped out and in as a whole.\\
Ideally, all elements of a single requested \verb+managedPtr<>+ will be needed in one step of a calculation. If not, Rambrain might end up having to swap in many excess bytes to use just one or two elements. Fortunately enough, the same argument applies for normal CPU cache locality and developers are used to developing for this consecutive, local access scheme. For a review of the term locality and further hints please see for example \cite{Denning,Chellappa2008}. Therefore, existing and highly optimised libraries are perfectly suited to be used together with Rambrain.
	\section{Architecture and Design}

	\begin{figure}[h]\centering
	{\begin{tikzpicture}[scale=.8]
\draw[rounded corners] (-2.5,-1) rectangle  (2.5,4);
\node at (0,-1) [below] {frontend};
\draw[rounded corners] (3.5,-1) rectangle  (8.5,4);
\node at (6,-1) [below] {backend};
  \node [draw,rounded corners](adh) at (0,0) {\verb+adhereTo<>+};
\node [draw,rounded corners](manp) at (0,3) {\verb+managedPtr<>+};
\node [draw,rounded corners,dashed](mmem) at (6,3) {\verb+managedMemory+};
\node [draw,rounded corners,dashed](mswap) at (6,0) {\verb+managedSwap+};
\draw (adh)->(manp);
\draw (manp)--(mmem);
\draw (mmem)--(mswap);
\node at (-0,3.6) [color=green!50!black] {type specific allocation};
\node at (0,-.65) [color=green!50!black] {ensures data locality};
\node at (6,3.6) [color=green!50!black] {swap strategy};
\node at (6,-.65) [color=green!50!black] {disk storage};
\end{tikzpicture}}
\caption{\label{fig:arch} {\bf Architecture of Rambrain:} Rambrain is divided into four major classes, each serving a distinct purpose. The classes in dashed boxes are abstract classes.}
	\end{figure}
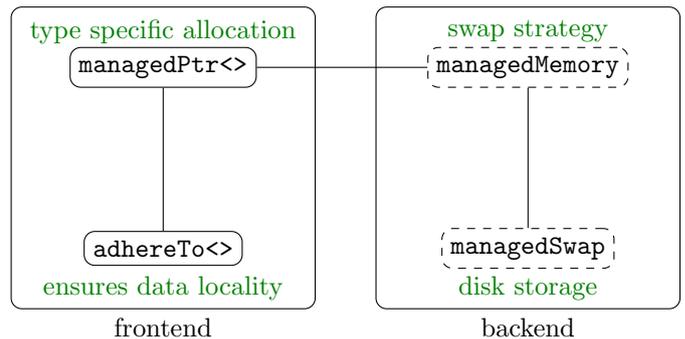
		\label{sec:arch}
Having described the interface of Rambrain, let us now describe how Rambrain is internally implemented and what design decisions have been taken to serve the user's data requests. As depicted in Fig. \ref{fig:arch}, Rambrain is divided into four independent classes.  While the user front end is implemented in a standardized way by the two classes \verb+managedPtr<>+ and \verb+adhereTo<>+, whose functioning has been described above, the abstract backend classes can be inherited to implement a custom strategy which elements to select for swapping. We currently serve two implementations of these classes each. One amounts to a dummy class that is used for testing purposes. The other implementations, \verb+cyclicManagedMemory+ as well as \verb+managedFileSwap+, will be described in the following sections. We provide profound source code documentation for all classes. The documentation can be compiled from source code using doxygen \citep{doxygen} or viewed online \citep{githubdoc,github} in a daily generated version.
		\subsection{Swapping Strategy}
		  \label{sec:archstrat}
It is a major design decision which elements to choose for swap-out to secondary storage when facing many currently not used objects. In this section we argue that a generic strategy should be at least capable of handling random access and access in the same order in an efficient way and describe the actual implementation.\\
When swapping out the same amount of data to media not capable of fast random access, swap-out size and fragmentation factors limit the speed achieved in a practical situation: The throughput per byte to be written/read is reduced when writing small chunks only, as the overhead of managing the transfer both physically and logically will take a greater fraction of execution time of the request. This is especially true when using hard disks as secondary storage: When fragments of the data needed are distributed over larger parts of the disk, the read/write head of the disk has to be positioned differently at every fragment. This process consumes more time than accessing consecutively stored data. While this argument does not apply for modern solid state disks any more, splitting data over multiple locations still poses an overhead as there must exist structures to describe and manage the splitting. Consequently a strategy writing out and reading in larger and consecutive parts at once will in general be faster than a strategy swapping out small chunks.\\
\begin{figure}\centering
\begin{tikzpicture}[scale=.75]

 \foreach \x in {1,2,3,4,5,6,7,8,9,10}{
   \draw ({cos((\x * 36.))*3.},{sin((\x * 36.))*3.}) arc({\x*36.}:{(\x+1)*36.}:3);
}
\foreach \x in {1,2,3,4,5,6,7,8,9,10}{
  \draw [fill=white]  ({sin((\x * 36.))*3.},{cos((\x * 36.))*3.}) circle(.25);
}
\draw [->,thick](0,2) --(0,2.5);
\node at (0,1.5) {active};
\node at (.75,-1.25) {counteractive};
\draw [->,thick] ({sin((4 * 36.))*2.},{cos((4 * 36.))*2.}) -- ({sin((4 * 36.))*2.5},{cos((4 * 36.))*2.5});
\draw [fill=red!75!black] (5,2) circle(.25) node [anchor=west,xshift=5] {swapped};
\draw [fill=green!75!black] (5,0) circle(.25) node [anchor=west,xshift=5] {allocated};
\draw [fill=blue!75!black] (5,-2) circle(.25) node [anchor=west,xshift=5] {pre-emptive};

\foreach \x in {5,...,7}{
  \draw [fill=red!75!black]  ({sin((\x * 36.))*3.},{cos((\x * 36.))*3.}) circle(.25);
}
\foreach \x in {0,...,4}{
  \draw [fill=green!75!black]  ({sin((\x * 36.))*3.},{cos((\x * 36.))*3.}) circle(.25);
}
\foreach \x in {8,9}{
  \draw [fill=blue!75!black]  ({sin((\x * 36.))*3.},{cos((\x * 36.))*3.}) circle(.25);
}

\draw [->] (0,3.5) arc (90:65:3.5);
\foreach \x in {0,...,9}{
  \node  [color=white]at ({sin(((9-\x) * 36.))*3.},{cos(((9-\x) * 36.))*3.}) {\x};
}

\end{tikzpicture}
\caption{\label{fig:cyclic} {\bf Cyclic managed memory:} Having accessed one element, it is very likely that the former next element will be the next one this time, too. Obeying this ordering, the algorithm will asynchronously pre-fetch ``pre-emptive" elements and swap out allocated but unused elements when necessary.}
\end{figure}
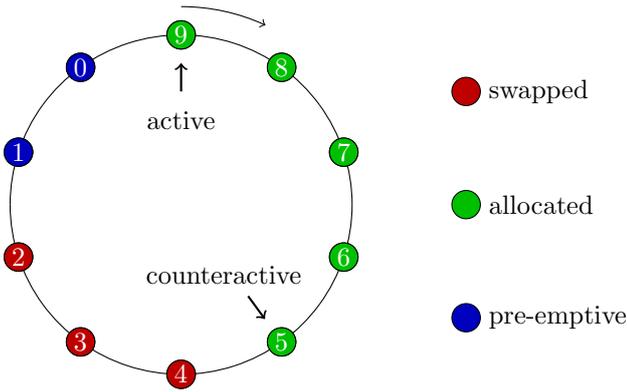
With no prior knowledge on what access pattern the user will impose on the data we can only make general assumptions and search for a strategy which can learn access patterns. The actual pattern encountered will lie somewhere in between the two extremes of a completely ordered and repeated sequence and random access patterns.\\
The Linux kernel for example tracks 'page age' and, when needed, preferably swaps out pages that have not recently been touched by the memory management subsystem. Without further going into details\footnote{The interested reader may consult e.g. \cite{linuxkernel} or https://linux-mm.org/}, this strategy has proven useful to general access patterns encountered on systems which have to swap memory occasionally. In the intended use case of Rambrain, however, the need to swap out data is an all present circumstance. Letting the user state which data is required currently, places Rambrain in a better situation than the kernel memory management is in. Rambrain is being actively told which data is not required any more and there exist hints, which data will be accessed by the application in near future. Thus, Rambrain can much more clearly specify the 'age' and 'ageing' of data in the application's context and also infer what to swap in next.\\
Thinking of looping over an array of data, which is very common in scientific codes, the most simple strategy is based on the assumption that if one element has been accessed right after the other, it repeatedly may be requested in that sequence in the future. Having accessed all elements, it is most likely that the first element will be accessed again. When there are multiple array objects, this also holds when a subset of objects is under consideration. Even when needing only a subset of all arrays, it is likely that the elements of the array will be accessed in the same order. This assumption suggests a cyclic strategy which we implement in the \verb+cyclicManagedMemory+ class and illustrate in Fig. \ref{fig:cyclic}. This order is represented as a doubly linked list of element pointers with connected end points.\\
To organize this as an effective queueing system, the most recently accessed element is marked with a so called ``active" pointer and the last still allocated and not swapped out element as ``counteractive". The counteractive element is followed by swapped out elements or elements that are in the process of being written to secondary storage. When accessed in an ordered way, we may keep elements in physical memory for as long as possible. The cycle defines a reasonable sequence of swap-out: the elements that have not been accessed for the longest time are the next candidates for swap-out. They are conveniently found by dereferencing the counteractive pointer and moving this pointer backwards as elements are swapped. This will write large chunks of data consecutive into the swap files. When a swapped out element is requested by the user, also the elements that are presumed to be needed next will be loaded pre-emptively and the elements will be placed in front of the former active element.\\
In this way, accessing the next element in a local sequence will be very fast as it can have already been loaded and no re-ordering has to be done to the cycle at all. Only the active pointer has to be moved backwards one element to apparently move all active elements one position forward in the cycle. As long as the arrays themselves will be accessed consecutively, local ordering is also preserved by this scheme when interchanging access to various arrays.

\subsection{Pre-emptive element swap-in and decay}
\label{sec:decay}
It is a non-trivial question to decide the amount of bytes which are to be swapped in pre-emptively. A pre-emptively swapped in element will use up free physical space. Thus one has to make sure to not load unneeded elements that would be swapped out again immediately. This could cause major increase of IO-operations, thereby slowing down the system. It is prevented by tracking the amount of pre-emptively swapped in bytes. Pre-emptive swap-in will take place only as long as only a certain number of pre-emptively loaded bytes or less are present. If a pre-emptively loaded memory element is accessed by the user, it's size will be subtracted from the pre-emptive budget. If an element has to be swapped in from the swap file, the next elements will be fetched too, until the pre-emptive budget is filled up again. In this way, random access does not cause additional overhead by swapping in unnecessary bytes as the pre-emptive budget will always be near its limit and thus no further pre-emptive elements are swapped in.\\
This procedure however can lead to a constantly filled up pre-emptive budget. Imagine that an array A fills the RAM completely before an array B is accessed consecutively. Given that some elements of A have been loaded pre-emptively, they will never be used while B is accessed. Thus, they effectively block the pre-emptive budget that would be useful in loading B consecutively. To avoid this situation, Rambrain implements a decay of pre-emptive elements. The amount of decaying pre-emptive elements is determined by probabilistic arguments to prevent random access from producing too many useless pre-emptive bytes in the following way:\\
The maximum size of the pre-emptive budget can be used to estimate the probability of hitting a pre-emptive element at random:\footnote{Assuming equally distributed element sizes which are only a fraction of the pre-emptive budget.} $$P_{\rm preemptive} \approx L_{\rm preemptive}/(L_{\rm ram}+L_{\rm swap})\leq L_{\rm preemptive}/L_{\rm ram}$$
Where $L_{\rm ram}$ is the maximum physical memory allowed, $L_{\rm swap}$ the amount of occupied swapped out bytes and \linebreak $L_{\rm preemptive}$ the size of the pre-emptive budget. Now, every time an element is not available in RAM, we determine the amount of pre-emptive elements that have been accessed since the last element had to be swapped in. The probability that these $N$ elements have been accessed randomly consequently can be estimated by $P_{\rm preemptive}^N$. If this value drops below 1 percent, we let decay twice the amount of the free pre-emptive budget, but at least one byte. Decaying implies swapping out pre-emptive elements to make space for new pre-emptive elements. This typically implies loading at least two elements pre-emptively, as the pre-emptive swap-in fraction is by default set to ten percent and this fraction squared equals the significance level assumed above.

\subsection{Swap file usage}
		  \label{sec:archswap}
When loaded into RAM, the data area of a \verb+managedPtr<>+ has to be allocated consecutively as pulling a pointer guarantees consecutive layout. On secondary storage devices we may split up the data over various swap file locations. While this is not desirable, it is of use when free swap file location is running out and we want to use smaller left-over chunks from previous deallocations.\\
Another major difference to managing heap memory, like the memory allocator in the standard libraries that is interfaced by the \verb+new+/\verb+delete+ operator implementations, is that one cannot easily use the free space for the management overhead. This is because the managing structures have to be accessible very fast and would cause considerable latency when resident in secondary storage.\\
Of course managing the chunks of the swap file in physical memory poses unavoidable overhead. It will limit the amount of managed memory as this overhead grows over the physical size of memory available. At the moment the user has to manage large enough data amounts in one \verb+managedPtr<>+ to keep this overhead small. While this sounds like reintroducing the problem we sought out to solve, we find a typical memory overhead to be 5 to 10 percent of the amount of allocated structures when the data content is about 1kB. This amounts to being able to manage half a terrabyte of data as if it were in RAM on a 32GB  machine. The data would be saved in roughly $5\cdot 10^{8}$ \verb+managedPtr<>+s of this size. It is advisable to switch to higher memory loads per \verb+managedPtr<>+ which reduces the overhead by the according factor, making more space addressable on disk. We plan to pack up objects into larger sets in future versions of the library to further reduce the overhead. It is also planned to monitor the overhead and strictly constrain it to the overall limit in future releases.\footnote{This, however, is a non-trivial task as typically the standard memory allocation implementation has the control over the system call extending heap size.}\\
Thus, given the task to swap out a \verb+managedPtr<>+, our standard implementation \verb+managedFileSwap+ checks its list of free chunks of memory in the swap files and tries to find the first free chunk the \verb+managedPtr<>+ fits into. If it fails to find such a chunk, it starts to split the data consecutively over the remaining gaps. If this also fails, it cleans up cached \verb+managedPtr<>+s produced by \verb+const+ accesses and tries again. If no free space is left, it will simply fail. As this unfortunate case may happen after days of calculation, we also provide a swap policy mechanism that states how the library should react in that case. Policies amount to ``fail in case of a full swap", ``ask the user if he wants to assign more swap space" or ``automatically extend swap space if free disk space is left to do so".

\subsection{Asynchronous IO and Direct Memory Access}
		  \label{sec:archaio}
\begin{figure*}[p]
	\centering
	\begin{tabular}{c}
		\scalebox{.85}{
			\begin{tikzpicture}
			\draw [dashed] (0,10)--(0,0);
			\draw [dashed] (5,10)--(5,0);
			\draw [dashed] (10,10)--(10,0);
			\node [draw,rounded corners,fill=white] at (-3,10) {object lifetimes};
			\draw [fill=white,dashed,path fading=north] (-.5,9) rectangle node[rotate=90] {}(.5,10) ;
			\node [draw,rounded corners,fill=white] at (0,10) {main thread};
			\node [draw,rounded corners,fill=white] at (5,10) {Rambrain libraries};
			\node [draw,rounded corners,fill=white] at (10,10) {Kernel Asynchronous IO};
			\draw [->,thick](1,9) -- node [midway,above]{adhereTo::adhereTo()} (4.25,9);
			\draw [->,thick](5.75,8.75) -- node [midway,above]{io\_submit()} (9.25,8.75);
			\draw [fill=white] (-4,0) rectangle node[rotate=90,path fading=none] {adhereTo glue;}(-3,9) ;
			\draw [fill=white] (-3,0) rectangle node[rotate=90] {double *ptr;} (-2,8.5) ;
			\draw [fill=white,dashed] (9.5,4.5) rectangle node[rotate=90] {asynchronous copy}(10.5,8.75) ;
			
			\draw [->,thick](1,8.5) -- node [midway,below]{request pointer} (4.25,8.5);
			\draw [dotted] (-3,9)--(0,9);
			\draw [dotted] (-3,8.5)--(0,8.5);
			
			\draw [fill=white,dashed] (4.5,8.75) rectangle (5.5,9) ;
			\draw [fill=white,dashed] (4.5,4.5) rectangle node [rotate=90]{waiting for IO} (5.5,8.5) ;
			\draw [<-,thick](5.75,4.5) -- node [midway,above]{io\_getevent()} (9.25,4.5);
			\draw [fill=white,dashed] (4.5,4) rectangle(5.5,4.5) ;
			\draw [<-,thick](1,4) -- node [midway,above]{returning pointer} (4.25,4);
			
			\draw [fill=white,dashed] (-.5,0) rectangle node[rotate=90] {actual work}(.5,4) ;
			\end{tikzpicture}}\\
		(a) Blocking IO\\
		\scalebox{.85}{
			\begin{tikzpicture}
			\draw [dashed] (0,10)--(0,0);
			\draw [dashed] (5,10)--(5,0);
			\draw [dashed] (10,10)--(10,4.25);
			\node [draw,rounded corners,fill=white] at (-3,10) {object lifetimes};
			\node [draw,rounded corners,fill=white] at (0,10) {main thread};
			\node [draw,rounded corners,fill=white] at (5,10) {Rambrain libraries};
			\node [draw,rounded corners,fill=white] at (10,10) {Kernel Asynchronous IO};
			
			\draw [->,thick](1,9) -- node [midway,above]{requesting ptr2} (4.25,9);
			\draw [->,thick](5.75,8.75) -- node [midway,above]{io\_submit()} (9.25,8.75);
			\draw [fill=white] (-4.5,0) rectangle node[rotate=90,path fading=none] {adhereTo glue2;}(-3.5,9) ;
			\draw [fill=white,path fading=north] (-2.5,4.5) rectangle(-1.5,9);
			\node [rotate=90] at (-2,6.75)  {adhereTo glue1;};
			\draw [fill=white] (-3.5,4.5) rectangle node[rotate=90] {double *ptr1;} (-2.5,8.5) ;
			\draw [fill=white,path fading=south] (-2.5,1) rectangle (-1.5,4.5);
			\node [rotate=90] at (-2,2.75) {adhereTo glue3;};
			\draw [fill=white] (-3.5,0) rectangle node[rotate=90] {double *ptr2;} (-2.5,4) ;
			\draw [fill=white,dashed] (9.5,4.75) rectangle node[rotate=90] {asynchronous copy}(10.5,8.75) ;
			
			\draw [<-,thick](1,8.5) -- node [midway,below]{returning ptr1} (4.25,8.5);
			\draw [dotted] (-3.5,9)--(0,9);
			\draw [dotted] (-3.5,8.5)--(0,8.5);
			\draw [dotted] (-3.5,0)--(0,0);
			\draw [dotted] (-1.5,4)--(0,4);
			\draw [dotted] (-1.5,4.5)--(0,4.5);
			
			\draw [fill=white,dashed] (4.5,8.75) rectangle (5.5,9) ;
			\draw [<-,thick](5.75,4.75) -- node [midway,above]{io\_getevent()} (9.25,4.75);
			\draw [fill=white,dashed] (4.5,4.25) rectangle(5.5,4.5) ;
			\draw [->,thick](1,4.5) -- node [midway,above]{requesting ptr3} (4.25,4.5);
			\draw [->,thick](5.75,4.25) -- node [below]{io\_submit()} (9.25,4.25);
			\draw [fill=white,dashed,path fading=south] (9.5,0) rectangle (10.5,4.25) ;
			\node [rotate=90] at (10,2.12)  {asynchronous copy};
			\draw [<-,thick](1,4) -- node [below]{returning ptr2} (4.25,4);
			
			\draw [fill=white,dashed] (-.5,4.5) rectangle node[rotate=90] {work on ptr1}(.5,8.5) ;
			\draw [fill=white,dashed] (-.5,0) rectangle node[rotate=90] {work on ptr2}(.5,4) ;
			\end{tikzpicture}
		}\\(b) Explicit asynchronous IO
	\end{tabular}
	\caption{\label{fig:sequence} {\bf Exemplary interaction of user code with Rambrain library.} Rambrain may be faster when giving clues about upcoming data requirements. While in (a) the time waiting for data to arrive is wasted, the user may use this idle time for calculations on already arrived data, as depicted in (b) and written in listing \ref{lst:exas}. As preventing idle time is highly desirable, Rambrain tries to behave like case (b) without the user explicitly hardcoding this. In order to do so Rambrain tries to guess the upcoming data demands of the program and automatically pre-fetches elements that will be needed.}
\end{figure*}
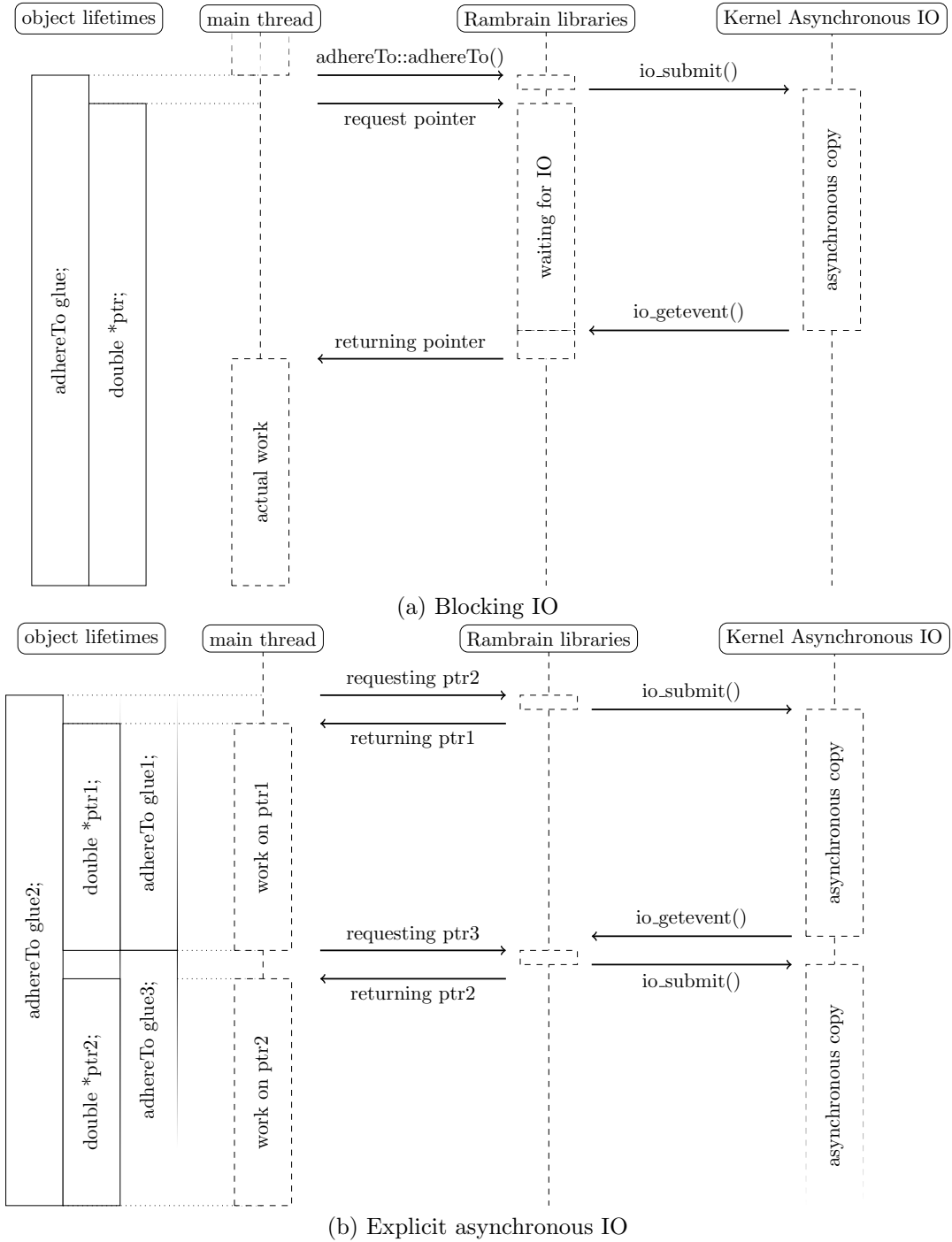

The main techniques to write out large data sets to secondary storage are Memory Mapping (MM), Direct Memory Access (DMA) and using Asynchronous IO (AIO) or a mixture of these. We briefly review the different approaches with respect to the task of transferring objects from primary to secondary storage:
\begin{itemize}
\item {\bf Memory Mapping}: The memory management unit in control of the virtual address space can be used to seemingly load contents of a whole file into physical memory. The same process used for paging will be utilised to write out or read in missing pieces and let an application use all space at once. When dealing with large files, this technique is very popular, as it is fast (may use DMA internally). However, when files become too big, the memory management unit quickly runs into similar problems to the one encountered with native swapping. A possible fix may be to map only parts of the swap files. In this case, however, one has to control tightly which mappings to close first, as closing will block when the mapped region is not written to disk completely. While there exists kernel hinting, a technique to tell the kernel which pages to write out first, the one-to-one mapping of allocations to the page file poses a bigger obstacle. Optimal decisions where to store certain elements are hard to find in a generic way and one is again limited to consecutive memory allocations. Splitting data would render pulling a pointer to consecutive memory impossible. Furthermore, the advantage of directly mapping allocations to swap file locations quickly can become a problem when the data has to be moved to still use a minimal memory mapped region. We thus quickly deferred using this method. There may be some interesting features to it, as automatic pre-fetching might already mimic an early stage of pre-emptive loading. Cleverly opening and closing such page-file ``windows", however, is hard to handle having no guarantees for future access patterns.
\item {\bf Direct Memory Access}: DMA can in principle copy parts of memory directly to secondary storage without routing the data through the CPU. It is fast in both throughput and latency. However, it imposes memory alignment restrictions on both sides and supports only writing chunks of a certain size (typically 512kB for hard disks). Since writing is direct, the action bypasses any buffering by the kernel and thus directly leads to disk access. While this can be advantageous in situations where one writes out many consecutive datasets and implements a write cache on ones own, it typically leads to overhead in our use case. Together with the imposed alignment restrictions, it is not clear how to write an efficient implementation without writing complex scheduling code or having lots of overhead when user objects do not fit into the DMA alignment. DMA, while fast, is very complex to handle in situations where a priori it is not clear what the user requests from Rambrain. Thus the benefits of fast IO and low CPU impact vanish in light of kernel file system buffering efficiency. There is a long going discussion involving Linus Torvalds who highly discourages the use of DMA by the user \citep[please see][]{linusrant}.
\item {\bf Asynchronous IO:} The Linux kernel provides the user with the possibility to asynchronously load and save data to file descriptors. Primary actions are taken only on the file system cache which has gone through a long evolution and is by now a very fast and efficient way to use free physical space without negative effects under high load. Furthermore, DMA or Memory Mapping techniques may be present in the background to bring the cache in sync with the secondary storage. Implementing Asynchronous IO upon normal buffering implies fast execution and efficient write-out while at the same time being robust to architecture changes. Finally the most efficient way of actually carrying out a certain storage operation may only be found out at system level.\\
The interested reader may be warned, however, that there currently exist three AIO implementations: kio (Kernel Asynchronous IO), libaio (which is just a C wrapper for the former) and POSIX AIO. The latter is currently implemented as blocking AIO, the former is not guaranteed to be truly asynchronous, as its implementation is file system driver specific. We use a pool of submitting threads using AIO to provide true AIO where possible and simulated AIO otherwise, using the libaio wrapper for the system calls. In this way, IO operations will be  non-blocking and have a low impact on CPU load.\\
By using asynchronous read and write requests, Rambrain is capable of loading data in the background with small impact on the CPU load. A technique for doing this is to first create the \verb+adhereTo<>+-object, which triggers swapping in of the object. While the asynchronous IO is swapping in the element, other calculations can be done. When finally pulling the requested pointer, it may already have been copied in in the background. A graphical scheme comparing synchronous and explicit asynchronous requests to Rambrain is available in Figure \ref{fig:sequence} and a schematic listing of the code producing this access scheme can be found in listing \ref{lst:exas}. Putting the highlighted line four after line six would constitute a synchronous version of the code. As the application can already process other data while fetching in next needed objects, this can effectively hide latency similar to GPU programming techniques or pre-fetching for caches \citep[see e.g.][]{prefetch}.\\
\begin{minipage}{\linewidth}\begin{lstlisting}[style=base,basicstyle=\small,label=lst:exas,frame=single,caption=Explicit asynchronous access]
managedPtr<double> data(1024),
                   data2(1024);
adhereTo<double> glue(data);
@adhereTo<double> glue2(data2);@
double* ptr = glue;
do_something_on_data(ptr);
double* ptr2 = glue2;
do_something_on_data2(ptr2);
\end{lstlisting}\end{minipage}
\end{itemize}
Having chosen AIO for transferring the data to secondary storage, the actual implementation is simple on the interface side but quite demanding on the scheduler side, as the scheduler has to deal with non-complete swap-outs and swap-ins when scheduling further action. As a rule of thumb, it has been found very useful to ``double-book" memory in the sense that chunks moving from or to physical memory will demand their size in both budgets. At the same time we also track the amount of memory which will be freed by such actions (and thus can be waited for when needed). When completed, the budget of free memory on the source side will be restored to the correct value and the bytes which were pending before will be subtracted from the pending bytes count. In this way, the scheduler can find the right strategy, given currently pending IO, and demand a small amount of IO to satisfy its constraints imposed by user requests.
\subsection{Compatibility to multithreading}
		\label{sec:multithreading}
Multithreading complicates writing the scheduler code a lot since one has to be very careful that the needs of one thread do not interfere with the needs of another thread. Scheduler and swap both are written as one instance shared by all local threads. This design decision was taken as data may be shared among threads and thus needs a common swapping procedure. Copying data between threads however will result in various \verb+managedPtr<>+s for each instance. This does not impose a big memory overhead since only the shallow control structures are possibly present multiple times and not the data themselves. Consequently, passing \verb+managedPtr<>+s and \verb+adhereTo<>+s from one to another thread has to happen thread-safely, as well as access to one \verb+managedPtr<>+ from multiple threads. Thread safety in this sense does not mean that one thread has exclusive access to a managed pointer, but that the mechanisms ensuring the availability of the data are written in a way that the object is present if at least one \verb+adhereTo<>+ in any thread is present and that the object may be swapped out at destruction of the very last \verb+adhereTo<>+ instance.\\
While reference counting is strongly related to the concept of shared memory parallelisation, a distributed memory setup is much easier described. Since every machine harbours it's own memory unit, it instantiates it's own management structures, swap and data pointers. Data are then copied between threads via the classical send and receive routines of the employed library, as for example MPI. This poses slightly more overhead than the shared memory case, but also provides the ability of a more intelligent access strategy especially if an asynchronous parallelisation model is implemented. One has to keep in mind, however, that if all machines or compute nodes write their swap files to the same disk, they may compete and slow down all IO, highly dependent on the timing of operations.\\
In total the amount of memory overhead due to parallelism should be negligible, especially since typical applications are globally memory dominated by the amount of data handled.

\section{Results and Discussion}
\label{sec:perf}
In this section we measure how code which utilises Rambrain compares to a code without Rambrain. Measuring performance is a non-trivial task for technical as well as theoretical reasons. First of all, tests should be reproducible and measure the overhead imposed by Rambrain. However, reaching this goal is non-trivial, as file system operations, kernel Asynchronous IO or scheduler performance in a multithreaded situation may affect the overall performance as well. Especially the typical use case - a developer seeking to work and debug on the same system - is hard to simulate in a reproducible and meaningful way. Separating library-imposed overhead and IO performance would be of no use either, as the user is interested in overall performance. Most of the carried out tests however will be highly speeded up by disk caching, which is also found in a productive system. We emphasize that while only RAM-to-RAM copying is done by the OS in these cases, these tests measure best the overhead implied by the workings and logic of the Rambrain library, since once the user is I/O limited, test results will be dominated by hardware performance.\\
In order to provoke swapping actions we set up a test system finding a PC with the smallest physical RAM module sizes removing all RAM modules up to one. The tests were then carried out using OpenSuse 13.2 (based on kernel 3.16) on an Intel(R) Core(TM)2 Quad CPU Q6700 operating at 2.66GHz on an ASUSTeK P5NT WS mainboard with 32Kb L1 Cache, 4MB L2 Cache and a standard unbranded 2GB memory module. The hard disk used is a Samsung SpinPoint S250.
		\subsection{Library overhead without swapping}
			\label{sec:nbody}
			
\begin{figure}
				\centering
				\scalebox{.75}{\input{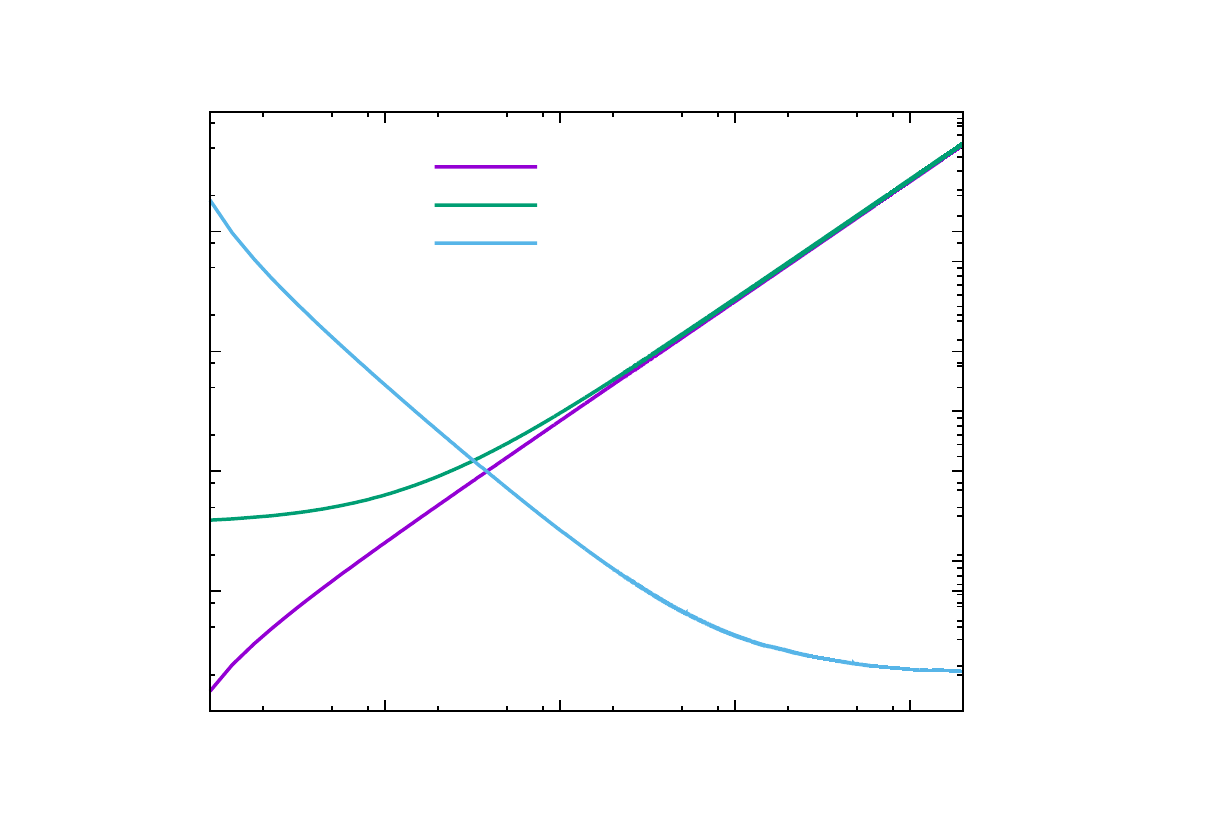}}
				\caption{\label{fig:nbody}{\bf Execution time of a n-body code:} We present timing information from a simple n-body code which accumulates data by saving particle trajectories and velocities. By comparing a version with and without rambrain we see that the overhead of the library amounts to only a few percent of execution time in the regime of reasonable data sizes.
					}
			\end{figure}

			We present the overhead the library imposes on the execution time of a user code in a regime, where actually nothing has to be swapped. This allows to judge whether Rambrain reaches near-to-native performance and thus can be employed if it is unclear whether it will be needed on the target system. We propose a test in which we perform a rather simple n-body simulation of a fixed set of particles using a Forward-Euler integrator \citep{1768ici..book.....E}. While each timestep only depends on the last position and velocities of all particles, we save the trajectories and velocities along the way in two dimensional arrays. A typical use case for this is in place visualisation of such a simulation. Therefore, the memory used by the program grows over time, adding two vectors per particle in each iteration. The results of both runs are shown in Fig. \ref{fig:nbody}.\\
			In the beginning of the simulation, when hardly any data is present, we notice quite a big relative overhead of the Rambrain library. However, this only amounts to an absolute difference of only one to two seconds. From a few MB of data on, both curves show the same scaling with time, which is given by the algorithm itself. The relative overhead presented by the blue line declines very rapidly and finally converges down to a value between one and two percent close to the two GB mark.\\
			In conclusion, a code utilising Rambrain is always a bit slower in the regime where no data has to be swapped out compared to native code. However, the impact on execution time is not a very big factor and we see no strict need for user to completely switch off Rambrain in this case.
			
		\subsection{Matrix operations}
			\label{sec:perfmat}
			
			In this subsection we demonstrate the internal movement of data for a common problem: Transposing a big matrix which itself does not completely fit in memory. We save matrices block wise, as it is done in many linear algebra libraries \citep[see e.g.][]{blas}. This allows for a straight forward migration to a Rambrain version of the algorithm, simply replacing one layer of pointers by a \verb+managedPtr<>+ class.\\
			\begin{figure}
				\centering
				\scalebox{.75}{\input{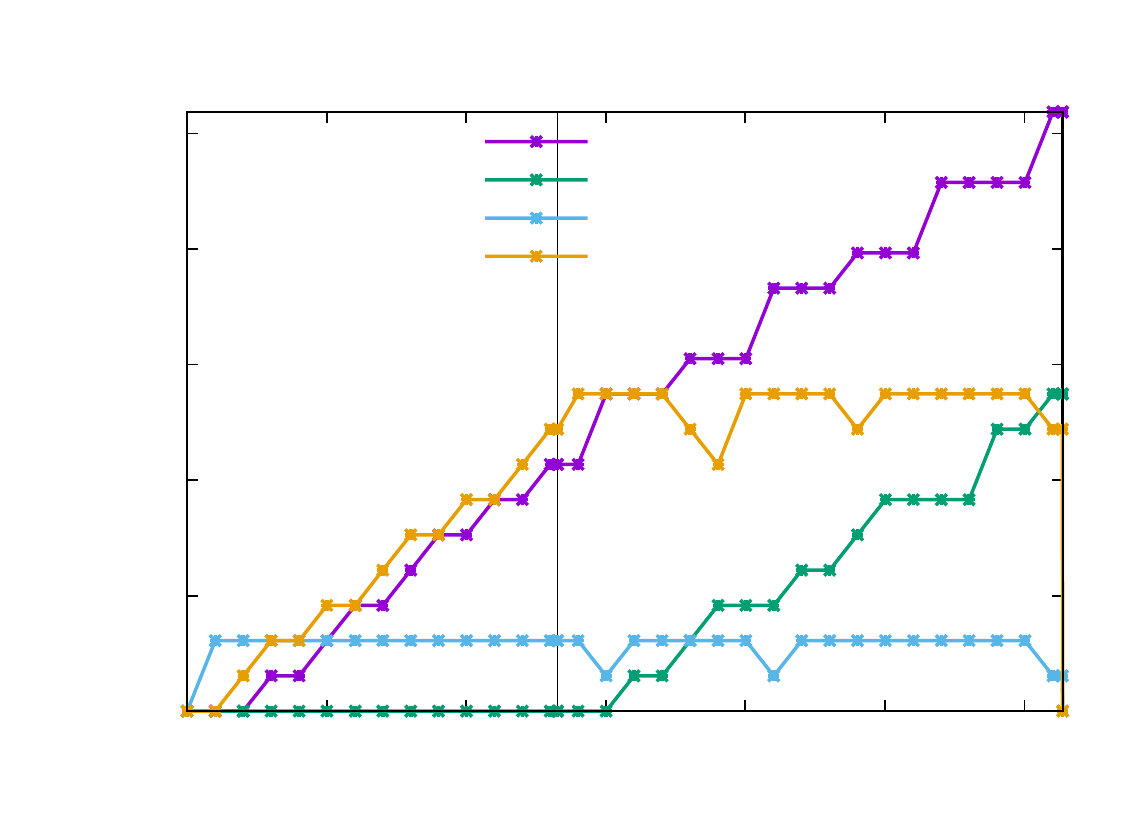}}
				\caption{{\bf Data movement for one 'Block' algorithm matrix transpose:} We show how data is moved between main memory and swap in one matrix transpose run. The vertical line marks the time point when the execution progresses from data allocation to the actual transposition.
					\label{fig:matrixtransposetiming}}
			\end{figure}
			The result is shown in Figure \ref{fig:matrixtransposetiming}. The left part of the plot shows the data allocation phase. At first the main memory is filled up very quickly with data, then data is consecutively swapped out to make room for more allocations. In the transposition phase afterwards, data is exchanged from swap to memory and back, loading all necessary blocks for the current transposition step. Please note that the asynchronous nature of Rambrain makes it very difficult to measure these values at a few discrete time points, since it is not clear when exactly the AIO events are handled in the background. Finally, the deletion of data is also plotted in the graph, but happens so fast that it is below the resolution limit of this plot. In total, we see that our design criteria are met and that Rambrain behaves well by constraining the usable memory. Additionally, the approximately linear scaling of the "Swapped Out" curve demonstrates, that the overhead of the library itself is not dependent on the current state of the memory.\\
			The diagnostic output leading to figures like this can be triggered directly in Rambrain, so that the user is able to easily profile the fundamental behaviour of his code.
			
		\subsection{Asynchronous IO and pre-emptive reading/writing}
			\label{sec:perfasync}
			
\begin{figure*}
				\centering
				\scalebox{.8}{\input{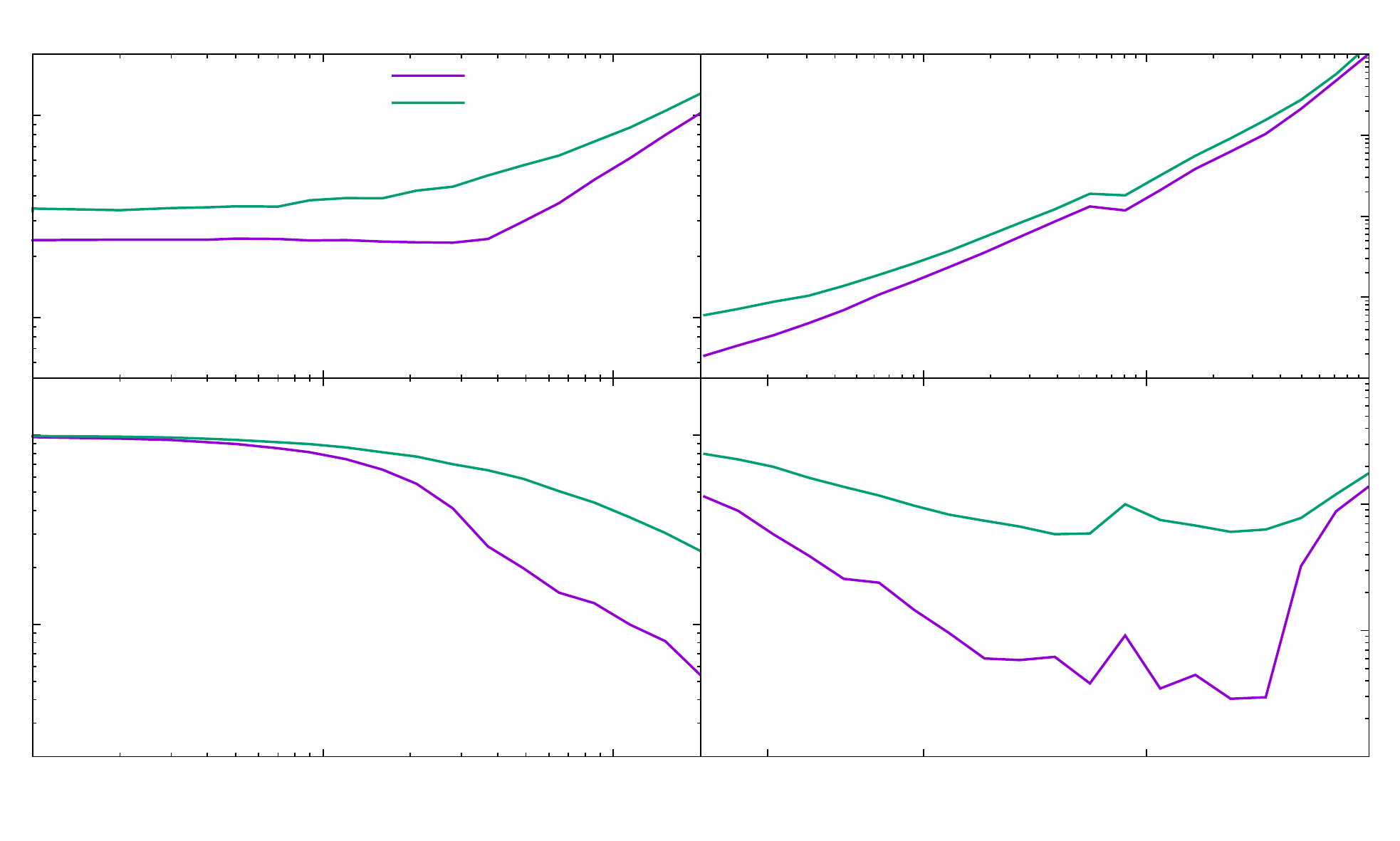}}
				\caption{{\bf Pre-emptive loading:} We compare enabled and disabled pre-emptive mechanism of Rambrain and find that the pre-emptive behaviour of Rambrain results in a significant performance boost.
					\label{fig:preemptiveasync}}
			\end{figure*}

			In this subsection we address the possible speed-up in execution time one can gain by efficiently using the asynchronous nature of Rambrain and the possibility to pre-emptively load and unload elements automatically.\\
			To measure the performance of this mechanism, we propose the test shown in listing \ref{lst:preempt}. We set up a two dimensional array which is realised by a list of managed pointers. While keeping the first dimension (i.e. the amount of one dimensional arrays) fixed at 1024, we vary the size of the underlying arrays (second dimension, $bytesize$). In order to measure the speed-up by asynchronism and pre-emptive actions we need to give the library some time to work in the background. Therefore, as in a typical use case, we iterate over the arrays in consecutive order and write the result of a simple integer multiplication into the respective array. We vary the percentage of the array that data is written to ($load$) and data chunk size, simulating an arbitrary computational load that scales with the data. The results of this test are presented in Figure \ref{fig:preemptiveasync}.\\
			\begin{lstlisting}[style=base,basicstyle=\small,label=lst:preempt,frame=single,caption=Standard implicitly asynchronous loading,float,floatplacement=H,xleftmargin=.6cm]
unsigned int numel = 1024, bytesize;
managedPtr<managedPtr<char>> arr ( numel, bytesize );
ADHERETOLOC( managedPtr<char>, arr, ptr );
float load;
float rewritetimes = load / 100.;
int iterations = 10230;

for ( int i = 0; i < iterations; ++i ) {
  unsigned int use = ( i % numel );
  //AdhereTo
  adhereTo<char> glue ( ptr[use] );
  //Pull the pointer to the object
  char *loc = glue;
  
  //Produce some computational load
  for ( int r = 0; r < rewritetimes * bytesize; r++ ) {
    loc[r % bytesize] = r * i;
  }
}
			\end{lstlisting}
			It is clearly observable that the execution time decreases due to pre-emptive strategy. Increasing the work which is done on the data in the left plot, the library's overhead is already masked at a few tens of percent of touched array elements. Working on the file buffer cache only, this test shows the minimal overhead of the Rambrain libraries. In a real use case scenario, the required computational load to completely mask swapping is increased. This result clearly encourages the user to leave the standard behaviour of pre-emptive support enabled whenever possible. Even if the data access is completely random, it does not imply a big performance drawback to try to be pre-emptive. Of course a problem-specific approach pre-fetching exactly the next needed elements without trying to guess can improve performance here. However, this strongly violates our assumption, that we value development time over execution time. We therefore argue that this optimisation leads towards developing a customized out-of-core algorithm, something no generic memory manager can substitute for. Be aware however, that when disk bandwidth becomes the limiting factor, only part of the swap in/out procedure can be masked by pre-emptive swap-in. For this reason, the overhead loading the data can become dominant when limited by the disk process and not carrying out enough calculations. While the pre-emptive strategy is still faster than not using the calculation time for loading in the next needed data in the background, the loading overhead in percent assimilates in the bandwidth-limited case. This can be seen in the lower right panel of the figure, as pre-emptive and non-pre-emptive strategies assimilate when disk-caching is not sufficient any more and write-outs to secondary storage dominate the timing.
		\subsection{Constant vs Non-Constant}
			\label{sec:const}
\begin{figure}[h]
				\centering
				\scalebox{.75}{\input{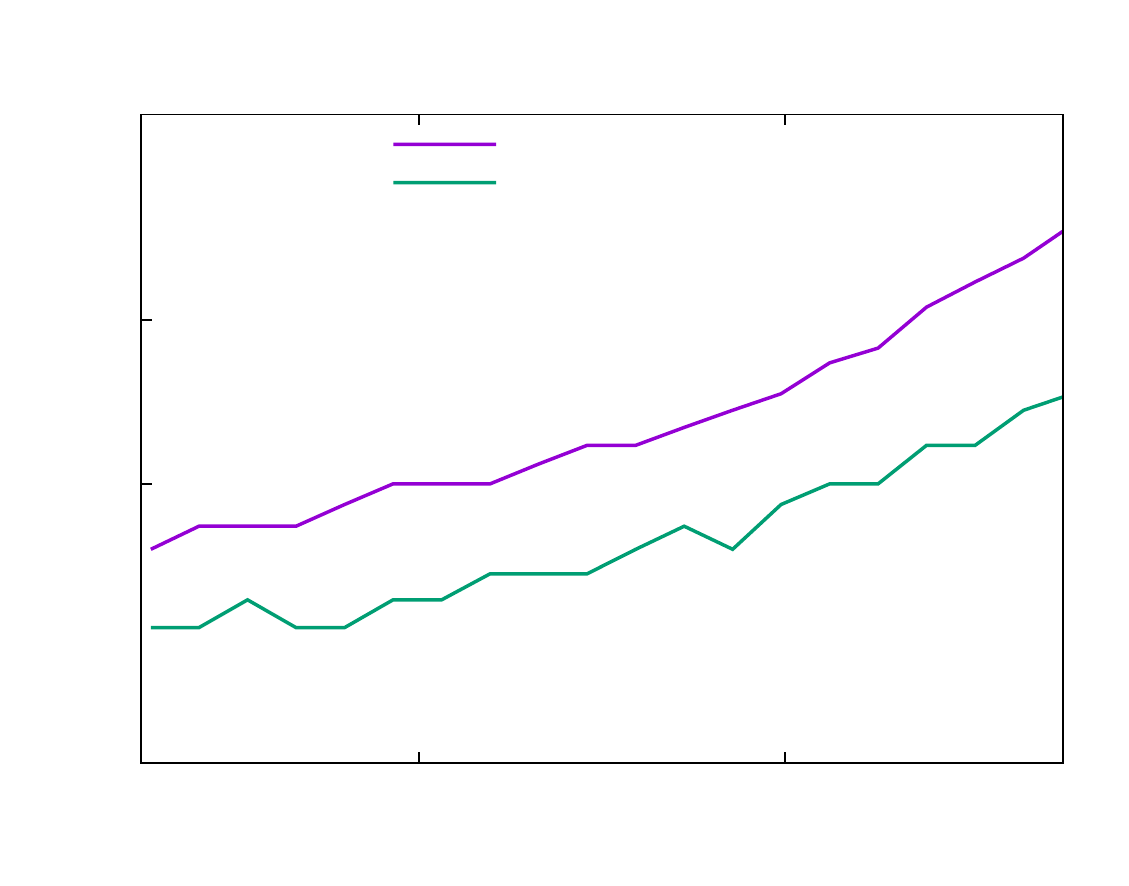}}
				\caption{{\bf Speed-up by pulling const pointers:} We run a simple test where data is drawn once as constant and once as writeable pointers and compare the time it takes to swap out the data afterwards in a regime where all the data still fits in the disk cache.
					\label{fig:compareconstspeedup}}
			\end{figure}

			Our next test is designed to examine how much time is saved by properly pulling \verb+const+ pointers when possible. As outlined in section \ref{sec:advacedusage} it is possible to request a pointer to constant data from an \verb+adhereTo<>+ object instead of a pointer to mutable data. This should be done in general, see e.g. \cite{meyers2012effective}, but is of special importance to the case of Rambrain. Not following this best practise will leave Rambrain with no clue on whether the data has been modified and forcing Rambrain to write the data out to the swap again. Hence, if the data has already a representation in the swap and is addressed as constant, this copy is kept as long as the swap has enough free space. When the in-memory copy of the data pointer is later deleted and a swap-out occurs, the data needs not to be written out again, saving expensive writing operations.\\
			In order to test this mechanism we allocate two blocks of data consisting of an array of smaller data chunks. The first one we call the real data while the second one is the dummy data which we will adhere to and pull a pointer from to ensure the real data being swapped out due to memory restrictions. Afterwards we access the real data and the dummy data in alternating sequence, once swapping in the data \verb+const+ and once non-\verb|const|. We measure the time it takes to swap in the dummy data in both cases, ergo capturing the time it takes to also swap out the real data. We present the resulting behaviour for different sizes of data blocks in Figure \ref{fig:compareconstspeedup}.\\
			We notice that the change in execution time by \verb+const+-access obviously scales with the amount of data, since it is highly dependent on the time it takes to complete the swap-out. In the regime of a data block amounting to between one and ten megabytes, we decrease the execution time of the relevant code sections by about 20 to 30 percent. Since these are relatively small data sizes in comparison to the main memory, we can assume that these data swap-outs are completely handled by the disk cache. Therefore we save only the time for cache management and basically a memory copy. When we enter the regime of secondary storage IO we can expect the difference in execution time to be even larger since the secondary storage itself is much slower than the main memory. For most storage types, storing data takes longer than reading data, thus we expect this mechanism to save even more time in this case. It is strongly advised to use \verb+const+-access whenever possible, also in light of other caches' properties and optimizations being used by the compiler.
			
	\subsection{Comparison with native OS swapping}
	Finally, let us compare the performance of Rambrain and system swapping. In principle, a local administrator can equip a Linux system with more swap space than usual by creating additional swap files or partitions with the system command \verb+mkswap+ and enable them for use with the command \verb+swapon+. However, please note that it is not possible to do so as a normal user. Additionally, this approach requires the allocation of the whole swap file space on secondary storage already in the beginning - regardless of how much of it will be actually used. Using this technique we create and enable a 10GB swap file on the described test machine.\\
	We compare a code which uses Rambrain to a non-managed code utilising system swapping. We carry out two different runs: In the first one, data is written consecutively to an 8GB sized matrix. In the second one, the application randomly writes to elements of this matrix. In the latter test we explicitly disabled the pre-emptive swapping algorithm.\\
	On some attempts to run unmanaged, the native application is killed by the OOM-killer. This probably happens due to the fast growth of heap memory. Also having a swap file which is not at least about 25 percent bigger than the actual swapped size often provokes the OOM-killer to terminate the process. Even if the OOM-killer does not kill the test process, it may be that it may shut down other processes in the background to free memory for the test process. When the attempts succeed, the system is virtually unusable as even opening another shell prompt takes minutes. Furthermore, the interference of the native code with the system does not stop when the application exits, but leaves the system in a slowly reacting state for minutes to hours of usage, as large parts of other applications and system processes have been swapped out to disk. We expect that running other applications such as an integrated development environment in parallel will aggravate the situation when trying to solve the problem using OS swapping.\\
	But also the actual execution time of Rambrain-managed code is favouring the use of our library. In the case of consecutive access, the version using Rambrain is about 10 percent faster than the native version. In case of random access, Rambrain is only 2\% faster than the native code, if we obey the design limitation that all elements of a single \verb+managedPtr<>+ will be accessed.\\
	This test result is further confirmed by daily experience of the authors being able to develop code on the same machine their analysis software runs in parallel without being disturbed by the process which uses Rambrain.
	
	\subsection{Real world application: Difference imaging}
		\label{sec:dia}
		
		To demonstrate that Rambrain is actually applicable to a real world problem, we choose a memory intensive difference imaging algorithm. The algorithm is designed to find variable light sources by comparison of multiple images. To mitigate errors due to noise these have to be convolved with a kernel first, before being subtracted from each other. For best accuracy, a variable point-spread function with a high number of free parameters is chosen as a kernel and an optimal version is computed by a minimisation technique. \cite{Alard2000} show that best results can be achieved by choosing one global kernel for the whole image. While this may seem to be the best approach anyway, we want to emphasize that usually only local kernels can be used because of the vast amount of memory consumption that arises in case of high resolution image material. This applies for example for the difference imaging code presented by \cite{Goessl2002} into which we embed rambrain in order to overcome exactly these limitations set by main memory.\\
		High resolution images taken with state of the art instruments \citep[see for example][]{Lee2012,Lee2015} can easily be of about $14000^2$ pixels in size, each. Typically kernels with several hundreds of free parameters are used which lead to an exemplary memory consumption by kernel matrices of
		\begin{align*}
			&ImageSize \cdot KernelSize \cdot (Values + Errors) \cdot float\\
			& = 14000^2 \cdot 400 \cdot 2 \cdot 4 B \approx 600 GB.
		\end{align*}
		This exceeds the physical size of main memory of a typical PC while the CPU time needed for such an analysis amounts to only a few hundred CPU hours.\\
		\begin{figure*}[th]
			\centering
			\includegraphics[width=\textwidth]{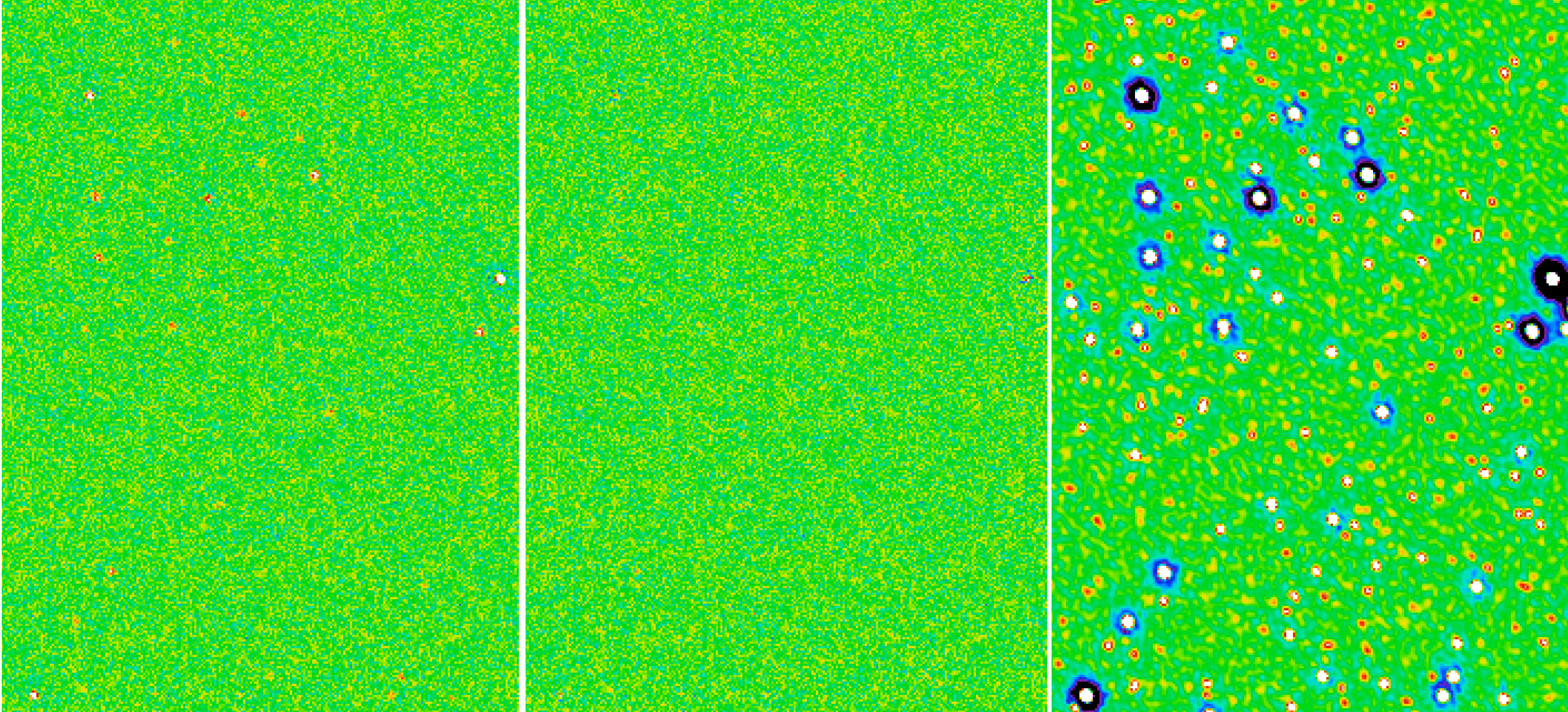}
			\caption{{\bf Difference imaging residual:} Left: Multiple local kernels; Middle: Global kernel with Rambrain; Right: Difference of both images. 
				\label{fig:dia}}
		\end{figure*}
		In Figure \ref{fig:dia} we present the results of such an analysis using simulated data. We assess the quality of the achieved fit by folding the reconstructed signal with the kernel and subtracting this from the input. If the kernel that has been constructed by the method reproduces the point spread function very well, the signal should vanish completely and only noise should remain. The left panel presents the result with a local kernel, where the image has been subdivided into several parts in order to fit into memory. The still present starlike features indicate that the kernel does not fit as well as in the middle image. This panel shows the global kernel in combination with rambrain and the right one displays the differences between those two. One can clearly see, that it does not only make a difference to use a global kernel, but being able to use this kind of global algorithm on the data leads to a result that contains a larger fraction of the signal of variable light sources in the sky.\\
		With Rambrain's capabilities to extend memory up to disk limitations, even more advanced algorithms can be applied without the typical memory restrictions. \cite{Barris2005} for example propose to use all unique pairs of images of a given set in order to calculate a yet more elaborate kernel. Memory management for this task can be delegated to a library suited, such as rambrain is. For further scientific analysis of actual observational images we refer to the upcoming paper of Riffeser et al. in prep.
		
	\section{Conclusions And Outlook}
		\label{sec:conclusions}
		We introduced the reader to writing code that utilises the Rambrain library. We described in detail why the proposed interface is sufficient to consistently handle data swap-out automatically and leads to satisfactory performance. We have demonstrated that the outlined mechanisms not only work properly, but also outperform naive approaches to mimic their strategy. Of course the library cannot compete with a fully specialized out-of-core algorithm, but can save a lot of development time in providing automatic facilities for large data sets. The library handles asynchronous transfer of data which provides latency hiding of disk IO operations and reduces idle times to a few percent if computational load allows. Furthermore, we have shown that the memory and CPU overhead of the library are both in the acceptable regime of only several percent. As all of this is provided by minimal user-side interaction, we feel the goal of writing a memory manager that enables the user to transparently access multiples of the physical memory to be fulfilled. As memory management is a short-cut to just stating what data is currently needed, the user can focus on the main goals of his application at the price of only a small overhead.\\
		We demonstrated the actual usage of our library via the example of difference imaging in astrophysics. However, the opportunities where such data intense problems rise to the surface of scientific work are vast and growing in numbers.\\
		The interested reader may find the code released as open source project \citep{github} accompanied by extensive further documentation, a list of the small set of prerequisites, notes about the (also system-wide) configuration options, a complete list of features and code examples. Interesting features are planned for future releases, such as direct mapping of file content to \verb+managedPtr<>+'s so that loading the data beforehand is not necessary any more.\\
		While currently the usage of Rambrain is only shown natively in other C++ codes, it is possible to interface and call the relevant functions also from codes written in different programming languages such as Fortran or Python. The library might lose some of it's elegance regarding the usage of strict scoping in C++, however we expect it to be fully functional when interfaced correctly. Writing such interfaces in proper manor is also part of future plans. Since the code is open source and available on github, the interested reader is happily invited to collaborate and assist in the development of such future features.\\
		Carrying out over 100 automatic tests partly consisting of random interaction with the library on every development step and keeping track of performance has proven very useful to find bugs which only occur under rare circumstances e.g. in multithreaded situations and improved robustness of the code a lot.\\
		We feel this library to be ready for use by a more general scientific audience.
	\section{Acknowledgements}
	We thank Karsten Wiesner and Christian Storm for helpful comments on presentation of advanced topics in this paper. We thank Arno Riffeser for working together with us on implementing and evaluating Rambrain in his difference imaging code. We thank Susanna Maurer for helping us organising a workshop on Rambrain at LMU. We also thank the anonymous reviewers, who helped to improve the overall quality of the paper and provided suggestions to complete the argumentation. Additionally we thank all others who helped us to find bugs in the actual implementation.
	\section*{Bibliography}
 	\bibliographystyle{apalike}
	\bibliography{rambrain}

\begin{thebibliography}{}

\bibitem[{Alard}, 2000]{Alard2000}
{Alard}, C. (2000).
\newblock {Image subtraction using a space-varying kernel}.
\newblock {\em Astronomy and Astrophysics, Supplement}, 144:363--370.

\bibitem[{Barris} et~al., 2005]{Barris2005}
{Barris}, B.~J., {Tonry}, J.~L., {Novicki}, M.~C., and {Wood-Vasey}, W.~M.
  (2005).
\newblock {The NN2 Flux Difference Method for Constructing Variable Object
  Light Curves}.
\newblock {\em The Astronomical Journal}, 130:2272--2277.

\bibitem[Blackford et~al., 2002]{blas}
Blackford, L.~S., Demmel, J., Dongarra, J., Duff, I., Hammarling, S., Henry,
  G., Heroux, M., Kaufman, L., Lumsdaine, A., Petitet, A., Pozo, R., Remington,
  K., and Whaley, R.~C. (2002).
\newblock {An Updated Set of Basic Linear Algebra Subprograms (BLAS)}.
\newblock {\em ACM Trans. Math. Softw.}, 28(2):135--151.

\bibitem[Callahan et~al., 1991]{prefetch}
Callahan, D., Kennedy, K., and Porterfield, A. (1991).
\newblock Software prefetching.
\newblock In {\em Proceedings of the Fourth International Conference on
  Architectural Support for Programming Languages and Operating Systems},
  ASPLOS IV, pages 40--52, New York, NY, USA. ACM.

\bibitem[Chellappa et~al., 2008]{Chellappa2008}
Chellappa, S., Franchetti, F., and P\"{u}schel, M. (2008).
\newblock {\em {Generative and Transformational Techniques in Software
  Engineering II: International Summer School, GTTSE 2007, Braga, Portugal,
  July 2-7, 2007. Revised Papers}}, chapter How to Write Fast Numerical Code: A
  Small Introduction, pages 196--259.
\newblock Springer Berlin Heidelberg, Berlin, Heidelberg.

\bibitem[Dementiev et~al., 2008]{SPE:SPE844}
Dementiev, R., Kettner, L., and Sanders, P. (2008).
\newblock {STXXL: standard template library for XXL data sets}.
\newblock {\em Software: Practice and Experience}, 38(6):589--637.

\bibitem[Denning, 2005]{Denning}
Denning, P.~J. (2005).
\newblock {The Locality Principle}.
\newblock {\em Commun. ACM}, 48(7):19--24.

\bibitem[Euler, 1768]{1768ici..book.....E}
Euler, L. (1768).
\newblock {\em {Institutionum calculi integralis}}.
\newblock Number Bd. 1 in {Institutionum calculi integralis}. imp. Acad. imp.
  Sa\`{e}nt.

\bibitem[{G{\"o}ssl} and {Riffeser}, 2002]{Goessl2002}
{G{\"o}ssl}, C.~A. and {Riffeser}, A. (2002).
\newblock {Image reduction pipeline for the detection of variable sources in
  highly crowded fields}.
\newblock {\em Astronomy and Astrophysics}, 381:1095--1109.

\bibitem[Imgrund and Arth, 2015]{github}
Imgrund, M. and Arth, A. (2015).
\newblock {\em {Github repository for Rambrain}}.
\newblock https://github.com/mimgrund/rambrain/.

\bibitem[Imgrund and Arth, 2017]{githubdoc}
Imgrund, M. and Arth, A. (2017).
\newblock {\em {Daily auto-generated documentation for Rambrain}}.
\newblock http://mimgrund.github.io/rambrain/.

\bibitem[{Imgrund} et~al., 2015]{imgrund2015}
{Imgrund}, M., {Champion}, D.~J., {Kramer}, M., and {Lesch}, H. (2015).
\newblock {A Bayesian method for pulsar template generation}.
\newblock {\em MNRAS}, 449:4162--4183.

\bibitem[{Lee} et~al., 2012]{Lee2012}
{Lee}, C.-H., {Riffeser}, A., {Koppenhoefer}, J., {Seitz}, S., {Bender}, R.,
  {Hopp}, U., {G{\"o}ssl}, C., {Saglia}, R.~P., {Snigula}, J., {Sweeney},
  W.~E., {Burgett}, W.~S., {Chambers}, K.~C., {Grav}, T., {Heasley}, J.~N.,
  {Hodapp}, K.~W., {Kaiser}, N., {Magnier}, E.~A., {Morgan}, J.~S., {Price},
  P.~A., {Stubbs}, C.~W., {Tonry}, J.~L., and {Wainscoat}, R.~J. (2012).
\newblock {PAndromeda - First Results from the High-cadence Monitoring of
  M31 with Pan-STARRS 1}.
\newblock {\em The Astronomical Journal}, 143:89.

\bibitem[{Lee} et~al., 2015]{Lee2015}
{Lee}, C.-H., {Riffeser}, A., {Seitz}, S., {Bender}, R., and {Koppenhoefer}, J.
  (2015).
\newblock {Microlensing Events from the 11 Year Observations of the Wendelstein
  Calar Alto Pixellensing Project}.
\newblock {\em The Astrophysical Journal}, 806:161.

\bibitem[Ligh et~al., 2014]{artofmem}
Ligh, M., Case, A., Levy, J., and Walters, A. (2014).
\newblock {\em {The Art Of Memory Forensics}}.
\newblock Wiley.

\bibitem[Meyers, 2012]{meyers2012effective}
Meyers, S. (2012).
\newblock {\em {Effective C++ Digital Collection: 140 Ways to Improve Your
  Programming}}.
\newblock Pearson Education.

\bibitem[Reiley and van~de Geijn, 1999]{Reiley1999}
Reiley, W.~C. and van~de Geijn, R.~A. (1999).
\newblock Pooclapack: Parallel out-of-core linear algebra package.
\newblock Technical report, Austin, TX, USA.

\bibitem[Rodrigues, 2009]{oom1}
Rodrigues, G. (2009).
\newblock {Taming the OOM killer}.
\newblock {\em LWN.net}.

\bibitem[Rusling, 1998]{linuxkernel}
Rusling, D.~A. (1998).

\bibitem[Salmon and Warren, 1997]{Salmon97}
Salmon, J.~K. and Warren, M.~S. (1997).
\newblock Parallel, out-of-core methods for n-body simulation.
\newblock In {\em PPSC}. SIAM.

\bibitem[Tang et~al., 2004]{Tang2004}
Tang, J., Fang, B., Hu, M., and Zhang, H. (2004).
\newblock A parallel out-of-core computing system using pvfs for linux
  clusters.
\newblock In {\em Proceedings of the International Workshop on Storage Network
  Architecture and Parallel I/Os}, SNAPI '04, pages 33--39, New York, NY, USA.
  ACM.

\bibitem[Toledo, 1999a]{Toledo}
Toledo, S. (1999a).
\newblock {A survey of out-of-core algorithms in numerical linear algebra}.
\newblock In Abello, J.~M. and Vitter, J.~S., editors, {\em {External Memory
  Algorithms}}, {DIMACS Series in Discrete Mathematics and Theoretical Computer
  Science}, pages 161--179. American Mathematical Society.

\bibitem[Toledo, 1999b]{Toledo1999b}
Toledo, S. (1999b).
\newblock External memory algorithms.
\newblock chapter A Survey of Out-of-core Algorithms in Numerical Linear
  Algebra, pages 161--179. American Mathematical Society, Boston, MA, USA.

\bibitem[Torvalds, 2002]{linusrant}
Torvalds, L. (2002).
\newblock {O\_DIRECT performance impact on 2.4.18}.
\newblock {\em Newsgroup fa.linux.kernel}.

\bibitem[van Heesch, 2015]{doxygen}
van Heesch, D. (2015).
\newblock {\em {Doxygen project webpage}}.
\newblock http://www.stack.nl/~dimitri/doxygen/index.html.

\bibitem[Vitter, 2001]{Vitter}
Vitter, J.~S. (2001).
\newblock {External Memory Algorithms and Data Structures: Dealing with Massive
  Data}.
\newblock {\em ACM Comput. Surv.}, 33(2):209--271.

\end{thebibliography}
	
\end{document}